# High-fidelity multimode fibre-based endoscopy for deep-brain *in vivo* imaging

Sergey Turtaev[1,5], Ivo T. Leite[1,4], Tristan Altwegg-Boussac[2], Janelle M.P. Pakan[2,6], Nathalie L. Rochefort[2,7*] and Tomáš Čižmár[1,3,4*]

[1]*Leibniz Institute of Photonic Technology, Albert-Einstein-Straße 9, Jena, 07745, Germany*
[2]*Centre for Discovery Brain Sciences, University of Edinburgh, Hugh Robson Building 15, George Square, Edinburgh, EH8 9XD, United Kingdom*
[3]*Institute of Scientific Instruments of CAS, Královopolská 147, Brno, 612 64, Czech Republic*
[4]*School of Science and Engineering, University of Dundee, Nethergate, Dundee, DD1 4HN, United Kingdom*
[5]*School of Life Sciences, University of Dundee, Nethergate, Dundee, DD1 4HN, United Kingdom*
[6]*Center for Behavioral Brain Sciences, Institute of Cognitive Neurology and Dementia Research, German Center for Neurodegenerative Diseases, Leipziger Straße 44, Haus 64, Magdeburg, 39120, Germany*
[7]*Simons Initiative for the Developing Brain, University of Edinburgh, Edinburgh, EH8 9XD, United Kingdom*

*Corresponding authors:*
\* These authors jointly supervised this work

n.rochefort@ed.ac.uk;  tomas.cizmar@leibniz-ipht.de

**Keywords:** *Digital holography, Fiber optics imaging, Optical microelectromechanical devices, Microendoscopy, Deep-brain imaging*

**Progress in neuroscience constantly relies on the development of new techniques to investigate the complex dynamics of neuronal networks. An ongoing challenge is to achieve minimally-invasive and high-resolution observations of neuronal activity *in vivo* inside deep brain areas. A perspective strategy is to utilise holographic control of light propagation in complex media, which allows converting a hair-thin multimode optical fibre into an ultra-narrow imaging tool. Compared to current endoscopes based on GRIN lenses or fibre bundles, this concept offers a footprint reduction exceeding an order of magnitude, together with a significant enhancement in resolution. We designed a compact and high-speed system for fluorescent imaging at the tip of a fibre, achieving micron-scale resolution across a 50 μm field of view, and yielding 7-kilopixel images at a rate of 3.5 frames/s. Furthermore, we demonstrate *in vivo* observations of cell bodies and processes of inhibitory neurons within deep layers of the visual cortex and hippocampus of anesthetised mice. This study forms the basis for several perspective techniques of modern microscopy to be delivered deep inside the tissue of**

**living animal models while causing minimal impact on its structural and functional properties.**

Optical systems have traditionally been understood as a series of components acting in a predefined and determinate manner. This notion is currently undergoing a steep transformation due to rapid advances in the technology and methods for spatial modulation of light. Reconfigurable elements such as computer-controlled holograms now facilitate the deployment of unusual, and frequently very complex, optical media with physical or functional properties that bring unique advantages relevant to biomedical applications[1].

Exploiting multimode fibres (MMFs) as ultra-narrow, minimally invasive endoscopes is a promising example of this conception[2–10], since it allows overcoming the trade-off between the size of the optical element and the attainable resolution[11]. The nature of light transport through MMFs leads to a fast phase decorrelation of the individual optical modes propagating in the fibre, transforming (or scrambling) the incident wavefront into a seemingly random speckle pattern. This is where adaptive optics offers a solution to overcome the degradation of optical signals propagating in randomising media[12–15]. A number of recently developed techniques in this domain currently enable the output light fields to be tailored into any desired distribution across the distal fibre facet or any arbitrarily remote plane[2–5]. The imaging concept of laser-scanning microscopy utilised in this study relies on the formation of diffraction-limited foci behind the distal end of a fibre, combined with image reconstruction from the fluorescence signals collected and guided back through the same fibre towards its proximal end[4,7].

Spatial modulation of light based on digital micromirror devices (DMDs) have recently opened-up a spectrum of opportunities in this domain, significantly increasing the modulation refresh-rate by several orders of magnitude when compared to well-established nematic liquid crystal based devices[16–19]. This allows scanning the focus formed behind the distal end of the MMF at rates of several tens of kHz, thus acquiring images at speeds approaching video-rates. Furthermore, it has also been shown that the quality of DMD-generated foci is significantly enhanced[20]. Building on the above mentioned advances, the focus of researchers is currently shifting towards first implementations in bio-medically relevant settings, including *in vivo* applications[21].

In the present work, we devise a compact and high-speed imaging system capable of resolving micron-sized neural cells in an anaesthetised animal model through, up to our knowledge, the most minimally-invasive endoscopic probe. Relying on high-performance holographic methods and carefully optimised design, the system is capable of artefact-free imaging of both neuronal soma and dendritic processes deep into brain tissues, with diffraction limited resolution. The resolution limits are dictated only by the numerical aperture (NA) of the fibre probe, and the efficient speckle background suppression approaches the theoretical limits[20].

Designed in a modular manner, the system consists of laser, calibration, beam-shaping and sample modules, as illustrated in Fig. 1b. A DMD in the core of the beam-shaping module is employed in the off-axis regime, which, despite the binary amplitude nature of spatial light modulation, is able to control the phase of the optical fields coupled into the fibre. Previously demonstrated principles (see Methods) allow analysing the system's response at the distal end of the fibre to a set of predefined optical fields generated by the DMD[3,4,20]. This response, recorded by the calibration module, characterises the light propagation through the whole optical path, from the DMD chip to a desired focal plane at or behind the distal end of the fibre and can be quantitatively expressed in the form of a transmission matrix (TM) of the system[14,15]. The TM inherently takes into account all the aberrations in the optical system and allows calculating phase modulations, which, when applied to the DMD, result in the formation of diffraction-limited foci at any desired location across the focal plane. The sequential exposure of a fluorescent sample to such a series of foci, while collecting the fluorescent signals propagating backwards through the fibre towards a bucket detector, is equivalent to raster-scanning laser microscopy.

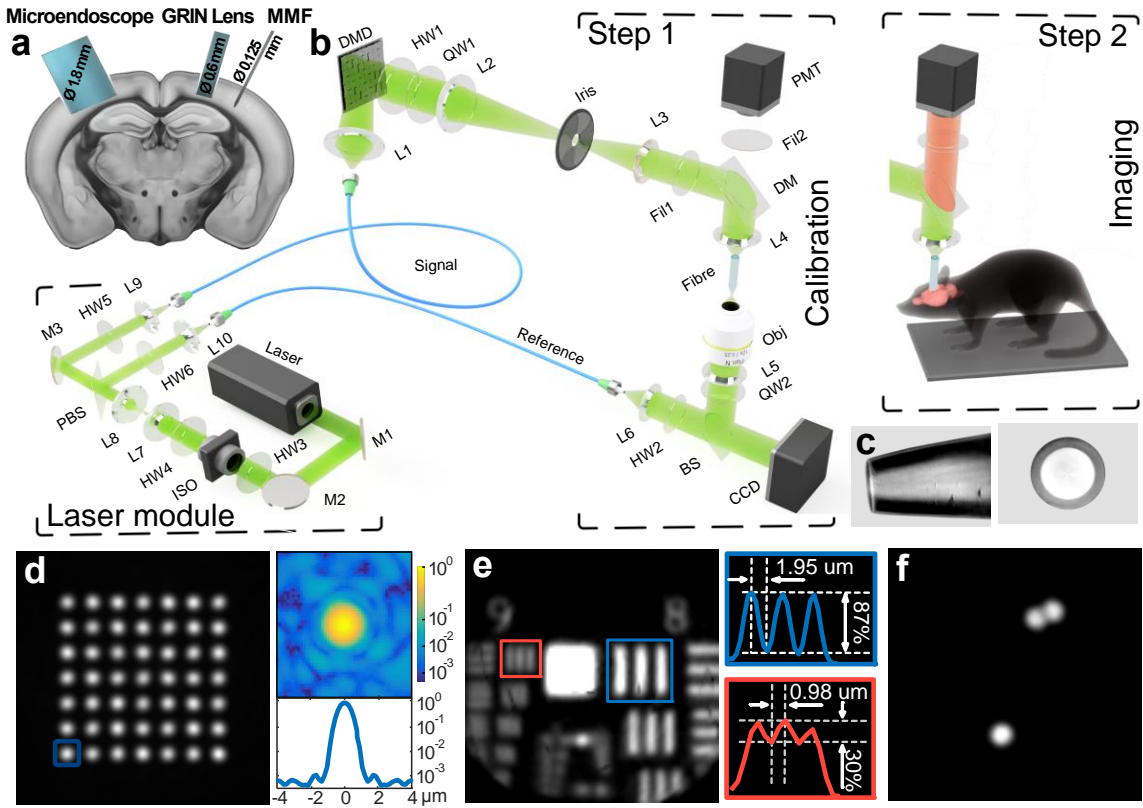

**Figure 1** Multimode fibre-based imaging system. (**a**) Scale-preserved comparison of the most typical endoscopic probes and a MMF. Image credit (brain slice): Allen Institute. (**b**) Scheme of the imaging setup. A calibration step precedes imaging. (**c**) Flat-top termination of the MMF probe with 50 µm core diameter and 60 µm external diameter. (**d**) Uniformity of foci generated across the fibre facet (left) and image and azimuthally-averaged profile of the chosen focal point in logarithimic scale (right). (**e**) Resolution assessment by a negative USAF-1951 test target. (**f**) Validation of fluorescent imaging using 4 µm fluorescent particles.

A standard commercially available MMF with a core of 50 µm in diameter and NA of 0.22 was chosen as the endoscopic probe. In order to minimise the tissue damage caused by the compression during the penetration process[22,23], the fibre probe (2 cm length) was post-processed into a flat-cone termination by polishing out the excess of cladding from 125 µm down to 60 µm external diameter, as depicted in Fig. 1c. For all experiments in this study, we offset the focal plane 5-15 µm from the distal end of the fibre, to guarantee that the full NA of the fibre was reached uniformly across the field of view and to minimise sample-induced aberrations. Once the probe is placed and aligned with the system, it takes approximately 2 min to obtain the TM, compute ~7000 phase modulation patterns for all desired foci across a 50 µm-wide circular region of the focal plane, and upload the patterns to the memory of the DMD device.

The fidelity of the synthesised foci is a crucially important attribute to reach the highest possible quality of imaging. In MMFs, as well as in all cases when light propagation through a randomising medium is controlled, only a fraction of the optical power leaving the medium can be directed into the diffraction-limited focus. The remaining optical power forms a background signal in the form of a speckle, observable in Fig. 1d. The speckled background does not affect the resulting image significantly when observing very sparse and high-contrast scenes (e.g. few fluorescent particles in the field of view), however, it manifests itself as a glare background reducing image contrast in the case of dense, volumetric samples.

The ratio between the power in the focal spot and the total output power emitted from the fibre is commonly used as a figure-of-merit when assessing the performance of a given optimisation approach. Working in the off-axis regime and utilising circularly polarised light has been shown to reach 75% of optical power stored in the desired focus, which is very close to the theoretical limit for the phase-only modulation based approaches[24]. Although more suited for turbid media where, in contrast to MMFs, the total transmitted power is not accessible, the enhancement is frequently employed as a quality metric. This parameter, defined as the ratio of the peak intensity on the focal point to the average level of speckled background, exceeded 3100 in our experiments.

Once the calibration procedure is finished, the calibration module is removed, and the system is ready for imaging. A fast bucket intensity detector being triggered by DMD reference TTL pulses (in a synchronous manner with switching between individual DMD modulations) allows operating the system with a refresh rate of 23 kHz, which results in an imaging rate of ~3.5 frames per second. The system has been devised with a robust cage-based construction, resulting in stable operation spanning several hours without the need for recalibration. The optical path was designed to achieve the maximum resolving power of the MMF. The resolution of the demonstrated fibre-based system has been assessed via a negative USAF 1951 test target placed in proximity to the distal end of the fibre. As shown in Fig. 1e, the separation of 4 µm and 1.9 µm between the lines could be imaged with contrasts of 85% and 30% respectively, which is in good agreement with the resolution limit (Rayleigh criterion) for the NA of the fibre and the wavelength used. For validation of operation in the fluorescent regime (Fig. 1f), we used 4.0 µm, red fluorescent beads that have been placed on a microscope slide. Emitted fluorescent light from the sample is collected and delivered back to the imaging system by the same fibre, where it is spectrally isolated from the excitation signal by a dichroic mirror and directed towards a photomultiplier tube.

For the demonstration of *in vivo* imaging capability deep within the brain of an anaesthetised animal, we used transgenic mice with a subpopulation of inhibitory interneurons labelled with a red fluorescent marker (tdTomato; see Methods). The insertion of the fibre endoscopic probe into the primary visual cortex (V1) and deeper into the hippocampus of the mouse brain was made through a small craniotomy. The images presented in Fig. 2a,b were recorded at different depths within V1 (0.5-0.8 mm) and in the CA1 region and dentate gyrus of the hippocampus (approximately 1.5 and 2 mm). The fibre probe causes minimal damage *in vivo*, as demonstrated by a post-mortem section of a perfused brain after fibre imaging (Fig. 2c), showing the impact of the insertion procedure with the fibre tract width not exceeding 50 µm. The resolution and contrast achieved *in vivo* allow for the visual identification of both relatively large objects, such as cell soma (diameter around 10-20 µm, Fig. 2a), and thin processes (usually 1-2 µm wide) with fine structures, corresponding to synaptic boutons (Fig. 2b). The high frame rate of the system allows live recording of the labelled structures during fibre penetration (see [Supplementary Movie](#)).

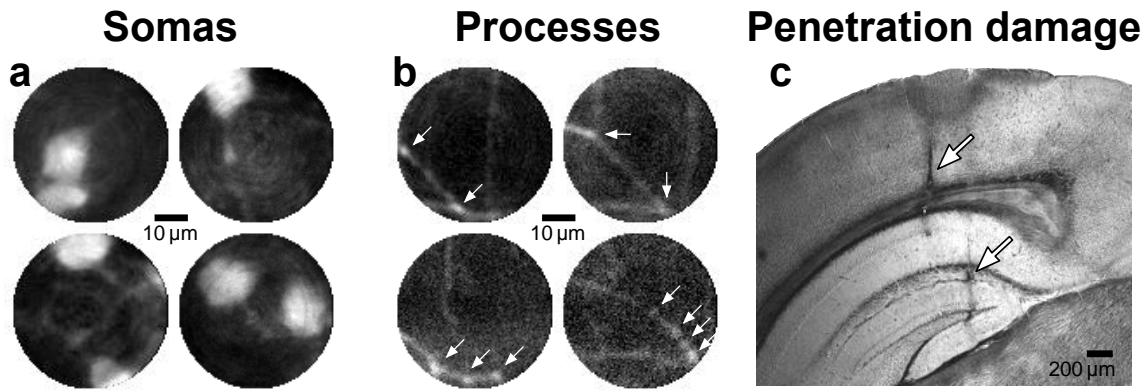

**Figure 2** *In vivo* implementation of MMF. Images of somata (**a**) and processes (**b**) of inhibitory neurons observed via direct insertion of the MMF probe up to 2-3 mm deep into a mouse brain. Arrows indicate branching points and synaptic boutons. (**c**) Post-mortem coronal brain section showing two fibre tracts of the MMF probe in the visual cortex and in the hippocampus. The width of each tract is less than 50 µm.

In this work, we have designed a highly optimised optical pathway for fluorescence-based imaging of deep-brain structures with micrometric spatial resolution and causing minimal damage to the tissue surrounding the fibre penetration area. Deploying the most efficient wavefront-shaping algorithms and, currently, the fastest possible hardware for light modulation, our system is capable of taking 7-kilopixel images with micron-size spatial resolution and imaging speeds of 3.5 frames per second allowing adequate spatial and temporal resolution for fluorescent imaging in living tissues. As a proof of concept, we were able to image fluorescent signal *in vivo* deep within a mouse brain through a single MMF-based probe. In anaesthetised transgenic mice with tdTomato fluorescent protein expression in a sparse subpopulation of inhibitory interneurons, we obtained visually identifiable images of neuronal somas and processes both in the visual cortex and in the hippocampus, more than 2 mm from the surface of the brain. While in our experiments the imaging was restricted to the cortex and hippocampus, with this technique the entire dorsal-ventral extent of the mouse brain (4-6 mm) can be covered, allowing imaging of even the most ventrally located nuclei in the brain. The achieved resolution is limited only by the numerical aperture of the fibre, being comparable with a standard 10× microscope objective. Imaging immediately after insertion of the endoscopic probe, which is only possible due to its minuscule footprint, eliminates the need for a long postoperative recovery period, as well as the necessity of surgical removal of overlying tissue and implantation of an imaging element, such as a GRIN lens[25,26]. This demonstration paves the way to *in vivo* implementation of numerous techniques of modern microscopy including multiphoton[9], super-resolution and light-sheet approaches[10], which will build further on the performance presented here. Future advancements will strongly rely on the development of new fibre types directly optimised for the purposes of holographic endoscopy[11].

## METHODS

**Setup**
As illustrated in Fig. 1b, the system is designed in a modular manner, consisting of laser, calibration, beam-shaping and sample modules. In the laser module, the beam is divided into a signal and a reference beam, which are coupled to single-mode polarisation-maintaining fibres. The calibration module is used for measuring the transmission matrix of the multimode fibre, and is subsequently replaced by a sample module for the anaesthetized animal model.

*Laser module.* A single-frequency, diode-pumped solid-state laser source emitting at the 532 nm wavelength provides a continuous-wave and linearly polarised beam. Achromatic doublets L7 and L8 form a telescope to demagnify the beam, and aspheric lenses L9 and L10 couple the signal and reference beams into the polarisation-maintaining fibres. Half-wave plate HW3, in combination with the input linear polariser of the Faraday optical isolator ISO, controls the combined power coupled to the signal and reference beams, whereas half-wave plate HW4, in combination with polarising beam splitter PBS, controls the power ratio between the signal and reference beams. Half-wave plates HW5 and HW6 are used to finely align the polarisation of the coupled beams with one of the birefringent axis of the polarisation-maintaining fibres (Fig. 1a).

*Beam-shaping module.* Achromatic doublet L1 collimates and expands the signal beam to overfill the DMD at an incidence angle of 24° with respect to the DMD chip. In this way, each micro-mirror in the 'on' state (+12°) reflects light in the direction of the optical system, whereas micro-mirrors in the 'off' state (-12°) redirect light towards a beam dump. Lenses L2, L3 and L4 relay the far-field of the first diffraction order of the holograms generated at the DMD to the proximal facet of the multimode fibre. An iris diaphragm is used in the Fourier plane of L2 to isolate the first diffraction order of the holograms, while blocking the remaining orders. The combination of lenses L2 and L3 is chosen to underfill the aperture of aspheric lens L4 in order to reduce its effective numerical aperture down to 0.23 matching that from the multimode fibre. The multimode fibre used has a core diameter of 50 µm and 0.22 NA, and at the 532 nm wavelength sustains approximately 2100 propagation-invariant modes (*i.e.* 1050 for each orthogonal polarisation state). Circular polarisation is well preserved after propagation through a straight fibre segment. The transmission matrix is measured between 3000 input modes and 7000 output modes (corresponding to the desired foci). Oversampling is essential to reach the highest fraction of the optical power contained in a focused spot. Dichroic mirror DM together with excitation filter Fil1 and emission filter Fil2 separate and spectrally purify the fluorescence signal. Photomultiplier tube PMT measures the overall intensity of the fluorescence signal. Half-wave plate HW1 and quarter-wave plate QW1 provide the two degrees of freedom necessary to reach the purest circular polarisation state of the spatially-modulated signal at the input facet of the fibre.

*Calibration module.* A movable, size-compact calibration module is used to acquire the TM of the fibre prior to the image acquisition. The TM is measured interferometrically using an external phase reference. Being uniformly distributed, the external reference prevents the formation of 'blind spots' originating from the speckled nature of internal references[3]. Microscope objective OBJ in combination with tube lens L5 image the selected focal plane (spatially offset away from the fibre facet) onto the CCD camera with 27.7× demagnification.

The reference beam is collimated by aspheric lens L6 and combined with the signal beam using non-polarising beamsplitter cube BS. Quarter-wave plate QW2 converts the polarisation state of the output speckle patterns to linear, and half-wave plate HW2 aligns the polarisation axis of the reference signal in order to maximise the signal-to-noise ratio of the interference patterns at the camera. The microscope objective is mounted on a single-axis translation stage for precise focusing and displacement of the calibration plane with respect to the output fibre endface.

**Calibration methods**
Binary amplitude gratings based on Lee hologram approach allow the DMD to be employed as a spatial light modulator in the off-axis regime[27]. The basis of input modes consists of truncated plane-waves of varying $k$-vectors. At the Fourier plane (focal plane of lens L2, Iris) as well as at the input fibre facet, this basis corresponds to a square grid of 65×65 focused spots. An alignment step prior to the TM acquisition measures the transmitted intensity by integrating the output speckle imaged by the CCD detector, for each input mode. This allows reducing the number of input modes to ~3000, since only those which are incident on the fibre core are effectively coupled. The basis of output modes consists of a square grid of 100×100 points spaced by approximately 0.5 µm across the focal plane (*i.e.* at a certain working distance from the output fibre facet), which are conjugate to pixels of the CCD detector during the calibration procedure. Only ~7000 output modes falling within Ø50 µm circular area are scanned during the image acquisition, at the maximum refresh rate of the DMD.

**Animals**
Data were acquired from 5 adult mice (5-6 months old). In 4 mice, a subpopulation of inhibitory neurons, Somatostatin-expressing (SST) neurons, was labelled with a red fluorescent marker (tdTomato) using a Cre-driver transgenic mouse line: Sst<tm2.1(cre)Zjh> (SST-Cre) [RRID:IMSR_JAX:013044] (Jackson Laboratory, ME, USA) cross-bred with Rosa-CAG-LSL-tdTomato [RRID:IMSR_JAX:007914] mice. In 1 mouse, another subpopulation of inhibitory neurons, (VIP) neurons, was labelled with the same red fluorescent marker (tdTomato) using a Cre-driver transgenic mouse line: Vip<tm1(cre)Zjh> (VIP-Cre) [RRID:IMSR_JAX:010908] (Jackson Laboratory, ME, USA) cross-bred with Rosa-CAG-LSL-tdTomato [RRID:IMSR_JAX:007914] mice. The animals were group housed (typically 2-4 mice), and both male and female mice were used for the experiments. All procedures were approved by the University of Edinburgh animal welfare committee and were performed under a UK Home Office project license.

**Surgical procedures**
For craniotomy, mice were anaesthetised with isoflurane (4% for induction and 1-2% maintenance during surgery and throughout imaging) and mounted on a stereotaxic frame (David Kopf Instruments, CA, USA). Eye cream was applied to protect the eyes (Bepanthen, Bayer, Germany), analgesics and anti-inflammatory drugs were injected subcutaneously (Vetergesic, buprenorphine, 0.1 mg/kg of body weight, carprofen, 0.15 mg, and dexamethasone, 2 µg). A section of scalp was removed, and the underlying bone was cleaned before a craniotomy (around 2×2 mm) was made over the left primary visual cortex (V1, 2.5 mm lateral and 0.5 mm anterior to lambda). Cyanoacrylate glue (Locktite, UK) was applied to the surrounding skull, muscle, and wound margins to prevent further bleeding.

*In vivo* **imaging**
The *in vivo* imaging stage was placed below the fibre and consisted of a custom made frame with ear-bars, to keep the animal's head position fixed during the imaging procedure, and a fitted facemask for delivery of isoflurane anaesthesia (1-2%). Suitable body temperature was maintained via thermal bandage. The imaging stage was mounted on a three-axis motorised translation stage with servo-driven actuators (ThorLabs, USA), allowing precise positioning of the animal both laterally for targeting the craniotomy and axially for the control of the fibre penetration process. The endoscopic fibre probe was gradually lowered into the craniotomy, up to 1-4 mm into the brain tissue targeting deep cortical layers in V1 and ventrally through the hippocampus to the base of the brain. Images were collected at ~3.5 frames per second at multiple regions throughout the tissue.

At the end of the imaging session, animals were given an overdose of sodium pentobarbital (240 mg/kg) prior to transcardial perfusion with phosphate buffered saline (PBS) and then 4% paraformaldehyde. The fixed brains were then extracted and 50 µm thick coronal sections were made with a vibratome (Leica, Germany) to confirm the location of the fibre tract.

**ACKNOWLEDGEMENTS**
This work was funded by: Marie Curie Actions of the European Union's FP7 program (608144 to S.T. and I.T.L., IEF 624461 to J.P. and MC-CIG 631770 to N.R.); European Regional Development Fund (ERDF: Center for Behavioral Brain Sciences to J.P. and CZ.02.1.01/0.0/0.0/15003/0000476 to T.Č.); European Research Council (ERC: 724530) to T.Č.; Thüringer Ministerium für Wirtschaft, Wissenschaft und Digitale Gesellschaft, Thüringer Aufbaubank and Federal Ministry of Education and Research, Germany (BMBF) to S.T., I.T.L. and T.Č.; Wellcome Trust, the Royal Society (Sir Henry Dale fellowship), Shirley Foundation, Patrick Wild Center, RS MacDonald Charitable Trust Seedcorn Grant, Simons Initiative for the Developing Brain to N.R.; University of Dundee and Scottish Universities Physics Alliance (PaLS initiative) to T.Č. Image credit of the brain slice (part of Fig. 1a): Allen Institute.

**CONFLICT OF INTERESTS**
The authors have no conflicts to declare.

**AUTHOR CONTRIBUTIONS**
S.T., I.T.L. and T.Č. designed the imaging system. S.T. developed all the software to run the system. T.A.B., J.P. and N.R. performed the *in vivo* experiments. S.T., T.A.B., I.T.L, T.Č. and N.R. analysed the results. N.R. and T.Č. led the project. All authors contributed equally to writing of the manuscript.